\documentclass[11pt]{article} 

\usepackage[utf8]{inputenc} 

\usepackage{geometry} 
\usepackage{graphicx} 

\usepackage[parfill]{parskip} 

\usepackage{booktabs} 
\usepackage{array} 
\usepackage{paralist} 
\usepackage{verbatim} 
\usepackage{subfig} 

\usepackage{fancyhdr} 
\usepackage{sectsty}
\allsectionsfont{\sffamily\mdseries\upshape} 

\usepackage[nottoc,notlof,notlot]{tocbibind} 
\usepackage[titles,subfigure]{tocloft} 

\usepackage{url}

\title{The Meaning of Probability of Default for Asset-Backed Loans}
\author{Graham Andersen\footnote{+61 438 696 600; graham.andersen@morgij.com.au}, David Chisholm\footnote{+61 438 562 359; david.chisholm@morgij.com.au\newline \newline \indent Morgij Pty Ltd \newline \indent L3, 10 Bond St \newline \indent Sydney Australia 2000}}

\date{28 June 2013} 

\begin{document}
\maketitle
\newpage
\section*{Abstract}
The authors examine the concept of probability of default for asset-backed loans.  In contrast to unsecured loans it is shown that probability of default can be defined as either a measure of the likelihood of the borrower failing to make required payments, or as the likelihood of an insufficiency of collateral value on foreclosure.  Assuming expected loss is identical under either definition, this implies a corresponding pair of definitions for loss given default.  Industry treatment of probability of default for asset-backed loans appears to inconsistently blend the two types of definition.

The authors develop a mathematical treatment of asset-backed loans which consistently applies each type of definition in a framework to produce the same expected loss and allows translation between the two frameworks.

\newpage
\section{Introduction}
Credit risk models generally analyse risk in terms of a Probability of Default (PD) and a Loss Given Default (LGD).  Both quantities (in this context) are predictive (as opposed to descriptive) statistical quantities representing future uncertain expectations of default and loss.  Specifically, PD is the probability that a given loan will go into default, and LGD is the likely loss (or expectation value of loss) that the lender will suffer if the loan goes into default.

The product of PD and LGD is then defined to be an Expected Loss (EL), which represents the expectation value of the lender’s loss, taking account of both the likelihood of the loan going into default and the likely loss should this occur\footnote{PD is by definition a probability.  LGD (and consequently EL) can be expressed as either a dollar value, or as a percentage of loan balance.  The distinction is not always important and we will clarify below where necessary.}.

Each of PD, LGD and EL can be defined at either the individual loan level or for a pool of loans.  Individual loan quantities can be aggregated to derive pool results.
Depending on how PD and LGD are defined and quantified, the resulting EL can serve a variety of purposes, including the following.

\begin{itemize}
\item EL can be identified as a minimum capital or credit support requirement, for example in credit rating or regulatory applications.
\item EL can be identified as an actual forecast loss for input into cashflow and pricing models.
\item EL can be used as an objective measure of risk for performance monitoring purposes.  For example a documented EL threshold may be used as a covenant by a lender.
\end{itemize}
In unsecured lending, the meanings of PD and LGD tend to be obvious and uncontroversial.  PD is simply the probability that the borrower becomes unable to fulfil their obligations from time to time (most importantly making any required principal and interest payments), and LGD is simply the shortfall between the outstanding loan balance and recoveries, should this occur.  Typically, expected recoveries are limited or even zero.  This simplicity is reflective of the fact that there is only one principal source of repayment, being the cashflows of the borrower.  While in practice a lender may have some security, say over the borrower’s assets generally, it is axiomatic to unsecured credit that the main perceived value of these assets is their ability to supply the borrower with cashflow which may be used to repay the debt, while little or no forward-looking credit is given to the liquidation value of these assets.

In asset-backed or secured lending, however, this becomes more complicated, since there are now two principal sources of repayment, the cashflows of the borrower \textit{and} the liquidation value of the security asset.  Generally, a lender relies on the borrower to make specified periodic principal and interest payments.  If the borrower fails to make such payments, then the lender may require the security asset to be sold and have the proceeds applied to the outstanding loan balance (including recovery costs).  In contrast to unsecured lending, it is generally axiomatic to secured lending that the asset has some significant liquidation value independently of the borrower's ability to make periodic repayments.

In this paper, we present a mathematical framework for the analysis of risk for asset-backed loans.

\subsection{Structure of Paper}
In the first section, we examine the meaning of probability of default for asset-backed loans, focusing on the practicalities of the default and recovery process.  We show that default and PD are capable of two very different meanings, one based on the borrower’s fulfilment of payment obligations and the other based on the sufficiency of the security asset value on recovery. We show that these imply two corresponding and different meanings for LGD.

In the second section, we show that existing credit risk model treatments of asset-backed loans appear not to treat PD, LGD and EL consistently, but mix the two definitions examined in the first section.

In the third section, we develop a consistent mathematical framework for analysing PD, LGD and EL for asset-backed loans.  The framework allows the choice of either definition of PD (and correspondingly LGD) as a basis for the quantification of EL, and further allows translation between the two definitions as the basis for the same quantification of EL.  The framework is based on treating LGD not as a single well-defined quantity (as is typical in existing models) but as a distribution based on the uncertainty in the value of the security asset.  EL is defined in this framework as a function of Loan-to-Value Ratio (LVR)\footnote{LVR is normally defined as the loan balance divided by the assessed value of the security asset value.  LVR as assessed on application is a fundamental risk metric in asset-backed lending.}.

In the fourth section we develop the inverse process of first taking EL as a function of LVR and then using our mathematical framework to derive the implied distribution of value of the security asset.

In the fifth section, we present some numerical examples illustrating the application of the developed framework.

\subsection{Unsecured and Asset-backed Loan Definitions}
For clarity, we define unsecured and asset-backed loans for the purposes of this paper.\footnote{Clearly intermediate situations are possible in practice, however this complexity is not important for this paper.}

An \textbf{Unsecured Loan} is a loan where the lender has no meaningful recourse beyond the cashflows of the borrower.

An \textbf{Asset-Backed Loan} is a loan where the lender has principal recourse to the cashflows of the borrower, but additional recourse to a specific asset which the lender has assessed as having a particular value.  In the event the borrower fails to meet their payment obligations under the loan, the lender may require the asset to be liquidated and the proceeds to be applied to repayment of the loan.

Arguably the clearest and most common example of an Asset-Backed Loan is a loan secured by a mortgage over residential property, with periodic principal and interest repayment requirements.  Accordingly, in this paper we will use such a loan as the generic example.

\section{What is ‘Default’ for an Asset-backed Loan}
\subsection{PD and LGD for Unsecured Loans}
In an Unsecured Loan, the borrower typically has an obligation to make certain payments of principal and interest at certain times.  An unremedied failure to do so is defined as an event of default and triggers the lender’s right to take recovery action.  Accordingly it is relatively simple to assess whether an Unsecured Loan has defaulted.  One simply determines whether the borrower has failed to make a required payment, and any allowed remedy period has passed without the appropriate remedy occurring.  Further, it is relatively simple to define a historical PD for a given sample of loans.  One simply determines whether or not each loan has defaulted as above and derives the proportion of loans which have defaulted.  This historical PD can then be used as an objective starting point for deciding on a forward-looking PD.

It is similarly simple to derive a historical LGD for the same sample.  One simply takes every loan which has defaulted (as defined above), and aggregates the ultimately unrecovered balance for each.  As for PD, this historical LGD can then be used as an objective starting point for deciding on a forward-looking LGD.

\subsection{Default for Asset-backed Loans}
The concept of default for an Asset-backed Loan is more complicated due to its dual recourse structure.

A typical sequence of events for a defaulting loan is as follows:
\begin{enumerate}
\item At some point after origination, the borrower stops making their required payments in full and on time.  We define this as “Initial Arrears”;
\item At some subsequent point, the borrower passes some defined arrears threshold and becomes sufficiently far in arrears that the lender is entitled to enforce their security and commence recovery action by liquidating the security asset.  We define this as passing an “Arrears Threshold”;
\item At some further subsequent point, the security property is sold and the proceeds are applied to the outstanding balance.  If the proceeds are sufficient to cover this balance, the lender suffers no loss.  If the proceeds are not sufficient, the lender suffers a loss\footnote{If a loss has been suffered, the lender may subsequently pursue the borrower for further recovery.  Typically such further recoveries are minimal or nil and do not cover any loss suffered.  We will not consider this eventuality in our analysis.}.  We define the process of selling the security property to repay the loan balance as “Liquidation”, whether or not the loan is in arrears.  Note that Liquidation can be carried out by either the lender or by the borrower, though in the latter case the lender typically monitors the process closely if the loan is in arrears.
\end{enumerate}

We can now ask which of these events represent default.

Few lenders would consider Initial Arrears to be the point where the loan defaults, as the borrower typically retains clear rights to cure the arrears, and a significant proportion of Initial Arrears do indeed cure.
Lenders do typically refer to passing the Arrears Threshold (commonly 90 days in arrears) as a default, as a high proportion of loans reaching this level of arrears proceed to Liquidation.

The complication arises with Liquidation, which event is specific to Asset-backed Loans.  If Liquidation results in a loss, then a lender would certainly agree that a default has occurred.  If no loss occurs, however, it is less clear that a lender would see that a default had occurred.  In particular, if Liquidation has been managed successfully by the borrower with the lender’s sanction and forbearance, the lender may, \textit{post hoc}, simply treat this as a normal sale, repayment and discharge.  It is relevant here that many mortgage loans are discharged early through sale of the property as a matter of course when the borrower moves residence, with no adverse credit behaviour.

Finally, we give some examples of unusual though possible situations which further complicate defining default:
\begin{itemize}
\item The borrower does not enter Initial Arrears, but voluntarily sells the security property for less than the outstanding loan balance (which we consider a form of Liquidation).  They repay part of the loan balance and declare bankruptcy, leaving the lender with a loss;
\item The borrower takes out a loan which allows capitalisation of interest and is never required to make a payment, subject to fulfilling some other covenant.  This covenant is breached and the security property is sold for a price which allows the loan balance to be repaid.
\end{itemize}
In the experience of the authors, the typical (though not universal) practical treatment from a credit analysis perspective is as follows:
\begin{itemize}
\item The lender’s loan servicing area describes loans in Initial Arrears and past the Arrears Threshold by reference to the number of days in arrears;
\item The lender’s capital management area describes loans in Initial Arrears as ‘in arrears’ and loans past the Arrears Threshold as ‘defaulted’;
\item When a lender provides information to external parties about historical defaults, they provide a record of all loans where a loss has occurred on Liquidation, but not on loans where the Arrears Threshold was breached but no loss occurred on Liquidation;
\item A lender will provide historical periodic data on arrears levels, including the level that corresponds to the Arrears Threshold, without distinguishing which of the loans eventually resulted in an actual loss.
\end{itemize}

In summary, then, it is clear that there are two different potential definitions of default:
\begin{itemize}
\item “Arrears Default” is where a borrower becomes unable to meet their periodic repayment obligations from their own cashflows and passes the Arrears Threshold;
\item “Liquidation Default” is where the proceeds of Liquidation of the security property are insufficient to cover the outstanding loan balance.
\end{itemize}

All other things being equal, it seems obvious that PD should generally be quantitatively different depending on the choice of default definition.  In particular, it seems intuitively obvious that the probability of Liquidation Default should depend strongly on LVR.  The higher the LVR, the lower the lender’s buffer against loss, so the higher the probability that a loss is incurred.  On the other hand, there seems no obvious reason why the probability of Arrears Default should generally depend on LVR as a primary variable, since LVR appears to have no fundamental linkage to the borrower’s cashflow position.\footnote{The qualification about LVR as a primary variable is critical.  The authors accept that Arrears Default may correlate with LVR as a secondary variable.  For example, serviceability (as measured by a ratio of the borrowers ability to pay divided by their obligation) will obviously correlate to loan size, which will obviously correlate to LVR.  The point is that a risk model ought to capture this through serviceability as an independent variable, so that a dependence on LVR would constitute ‘double-counting’ the risk.  As noted at the end of this paper, we also accept that there may be specific circumstances under which a dependence on LVR may be introduced.}

Given that EL should be quantitatively the same under both definitions (since we are assuming the same sequence of events), each version of PD will have a corresponding version of LGD.

\section{Existing Treatments of PD, LGD and EL}
The main publicly disclosed detailed models for analysing risk based on PD, LGD and EL for Asset-Backed Loans are those of the major credit rating agencies (“CRAs”) as disclosed in their various ratings criteria:
\begin{itemize}
\item Standard \& Poor's \cite{SP};
\item Moody's \cite{Moodys}; and
\item Fitch \cite{Fitch}.
\end{itemize}

In the authors’ experience, most credit risk models for Asset-Backed Loans used by market participants such as investors follow the same basic structure as, and are quantitatively similar to, the CRA models, so the CRA models can be taken as broadly representative of the overall market approach.
Accordingly, we examine below how the CRA models analyse Asset-Backed Loans, based on their current published rating criteria for residential mortgage-backed securities.  For the purposes of this paper, we will only examine the CRA approaches as far as developing their ‘base’ EL, PD and LGD as functions of LVR.  In particular we will not consider their treatments of:
\begin{itemize}
\item Adjustments reflecting particular characteristics of borrowers, loans or security properties;
\item Adjustments reflecting foreclosure costs;
\item Pool-level adjustments for things such as loan concentrations or the identity of the lender;
\item Adjustments reflecting loan performance; or
\item Adjustments reflecting cashflow analysis.
\end{itemize}

The authors acknowledge the above effects are important elements of credit analysis generally and of credit ratings in particular, however they are not relevant to the purposes of this paper, which focuses on the development of a ‘base’ treatment.
\subsection{PD}
\begin{figure}[htpb]
\caption{}
\includegraphics[width=120mm]{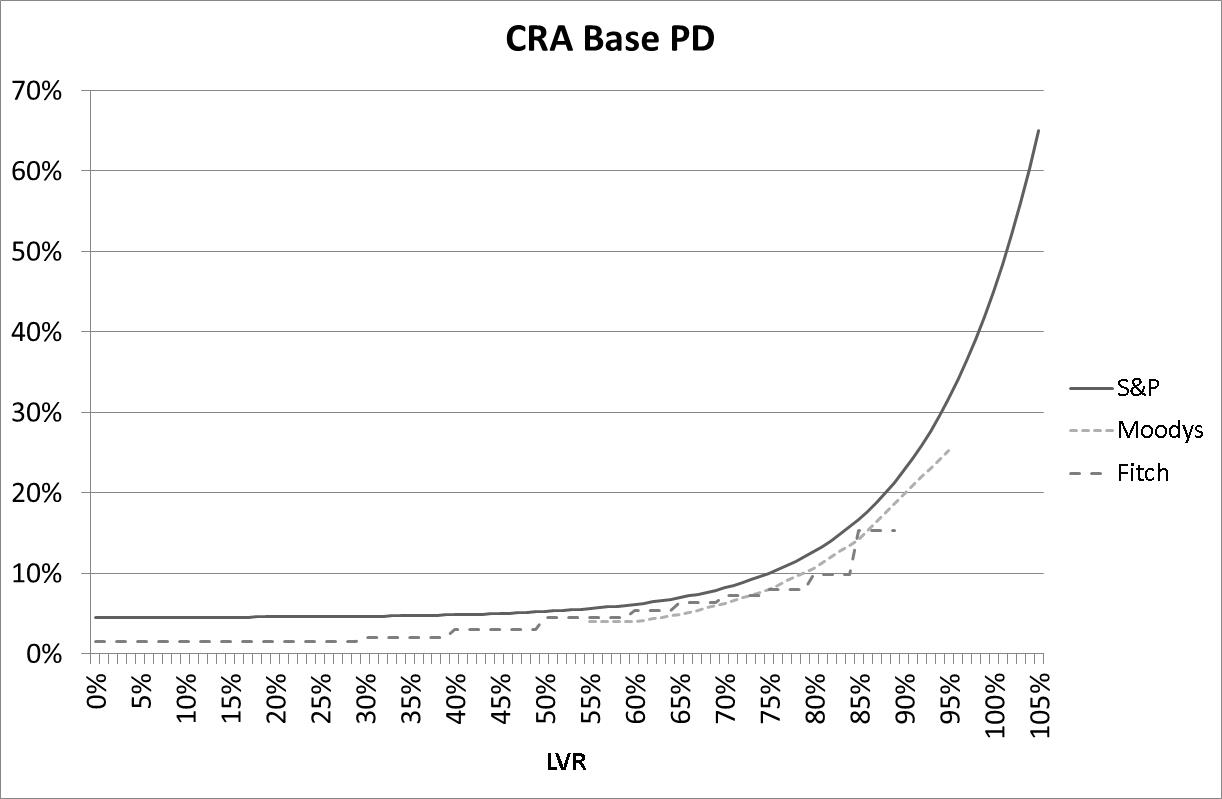}
\label{CRAPDChart}
\end{figure}

Standard \& Poor's and Fitch refer to ‘foreclosure frequency’ as their measure of the likelihood of default, while Moody's refers to ‘default frequency’.  In each case, it is clear from usage that the measure corresponds to a probability of default, and for simplicity we will refer to each as PD.
Figure \ref{CRAPDChart} shows base PD for rating levels AAA (Standard \& Poor's), Aaa (Moody's) and AAA (Fitch), based on each CRA’s current published criteria.
\begin{itemize}
\item In the case of Standard \& Poor's, base PD represents a benchmark PD of 10\% multiplied by a defined exponential function adjusting the benchmark PD for LVR.  We have terminated the LVR range at 110\%, where the base PD approaches 100\%;
\item In the case of Moody's, base PD represents tabulated PD at various LVR, interpolated for intermediate LVR.  Moody’s does not explicitly state that PD should be interpolated, however the accompanying chart in their criteria suggests this is the case.  We have only shown the range of LVR where Moodys publishes its base PD;
\item In the case of Fitch, base PD represents tabulated PD at various LVR.  Fitch specifies  PD in LVR bands, without interpolation.  We have terminates the LVR range at 95\%, where Fitch indicates that PD is determined on a case-by-case basis.
\end{itemize}

The different PD regimes appear broadly consistent with each other.
In each case PD is strongly dependent on LVR, which clearly implies the use of a Liquidation Default definition of PD.  In particular, we note that each curve has not only positive slope, but generally positive curvature, so that the increase in PD increases with increasing LVR.

\subsection{LGD}
Each of the CRAs refer to ‘loss severity’ as their measure of the likely loss in the event of a default.  In each case, it is clear from usage that ‘loss severity’ corresponds to a loss given default, and for simplicity we will refer to it as LGD.

Each of the CRAs specify a base Market Value Decline (“MVD”) as the expected percentage reduction in the value of the security property from its assessed value to the actual sale proceeds on Liquidation.  LGD is then calculated as the shortfall (if any) between the outstanding loan balance and the resulting sale proceeds on liquidation\footnote{The rating agencies increase the resulting LGD with an allocation for foreclosure costs, however we ignore this complication for the purposes of our analysis.  While these costs can be numerically significant, they can be applied to any framework as a simple adjustment to the ‘base’ LGD.}.  On this basis, we have:

\begin{equation}
LGD(L,M)=max\left[LDG_{min},\frac{(L-M-1)}{L}\right]
\end{equation}

\indent where

\begin{figure}[htpb]
\caption{}
\includegraphics[width=120mm]{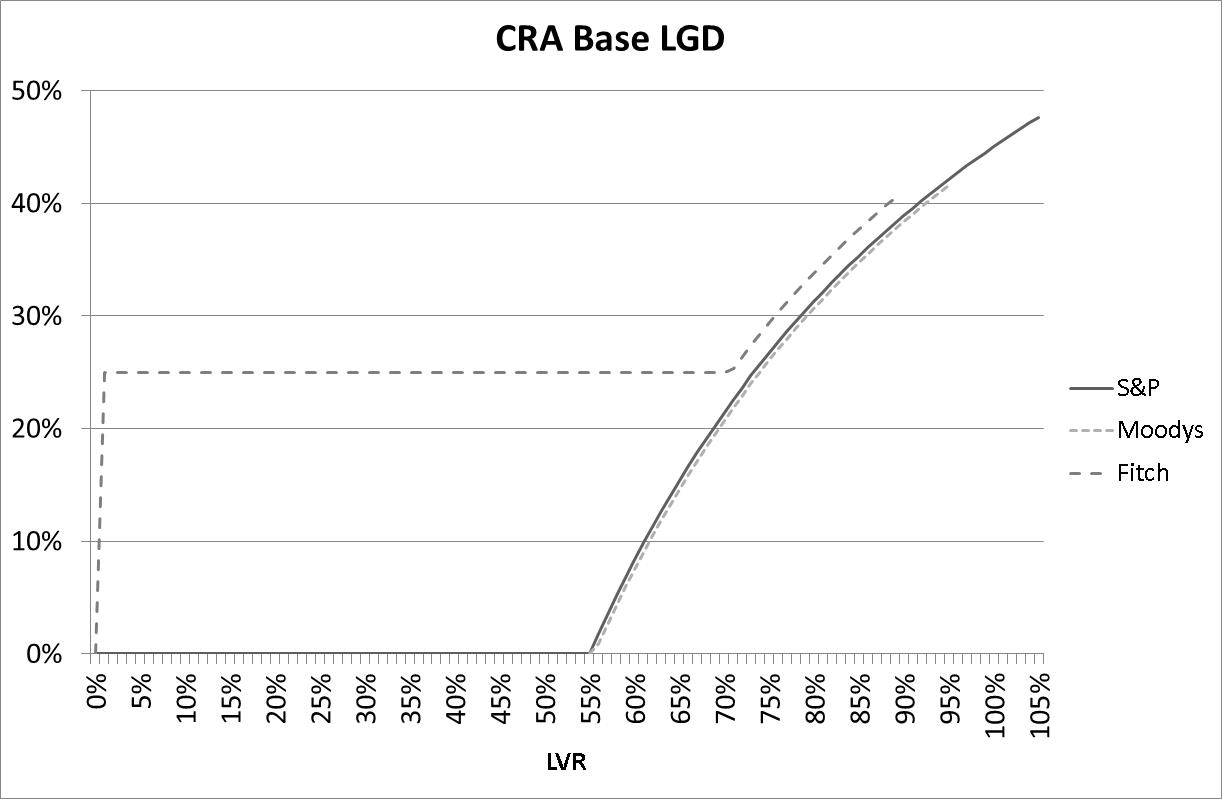}
\label{CRALGD}
\end{figure}

\begin{quote}
L is the LVR\footnote{We note that LVR must be non-zero.  In practice this is simply equivalent to assuming that the loan has a positive balance and the value of the security property is non-negative.}
\newline $M$ is the MVD, with the sign convention that negative M represents a reduction in property value
\newline $LGD_{min}$ is a defined minimum LGD
\newline $LGD(L,M)$ is the LGD as a function of LVR and MVD
\end{quote}
Figure \ref{CRALGD} shows base LGD based on each CRA’s current published criteria, and the same rating levels and LVR ranges specified above.
\begin{itemize}
\item In the case of Standard \& Poor's, base LGD is based on a MVD of 45\% and minimum LGD of 0\%;
\item In the case of Moody's, base LGD is based on  MVD of 44.5\% and minimum LGD of 0\%;
\item In the case of Fitch, base LGD is based on  MVD of 47\% and minimum LGD of 25\%.
\end{itemize}

As for PD, the different regimes appear broadly consistent, the main difference being Fitch’s adoption of a non-zero minimum LGD.
\subsection{EL Calculation}
\begin{figure}[htpb]
\caption{}
\includegraphics[width=120mm]{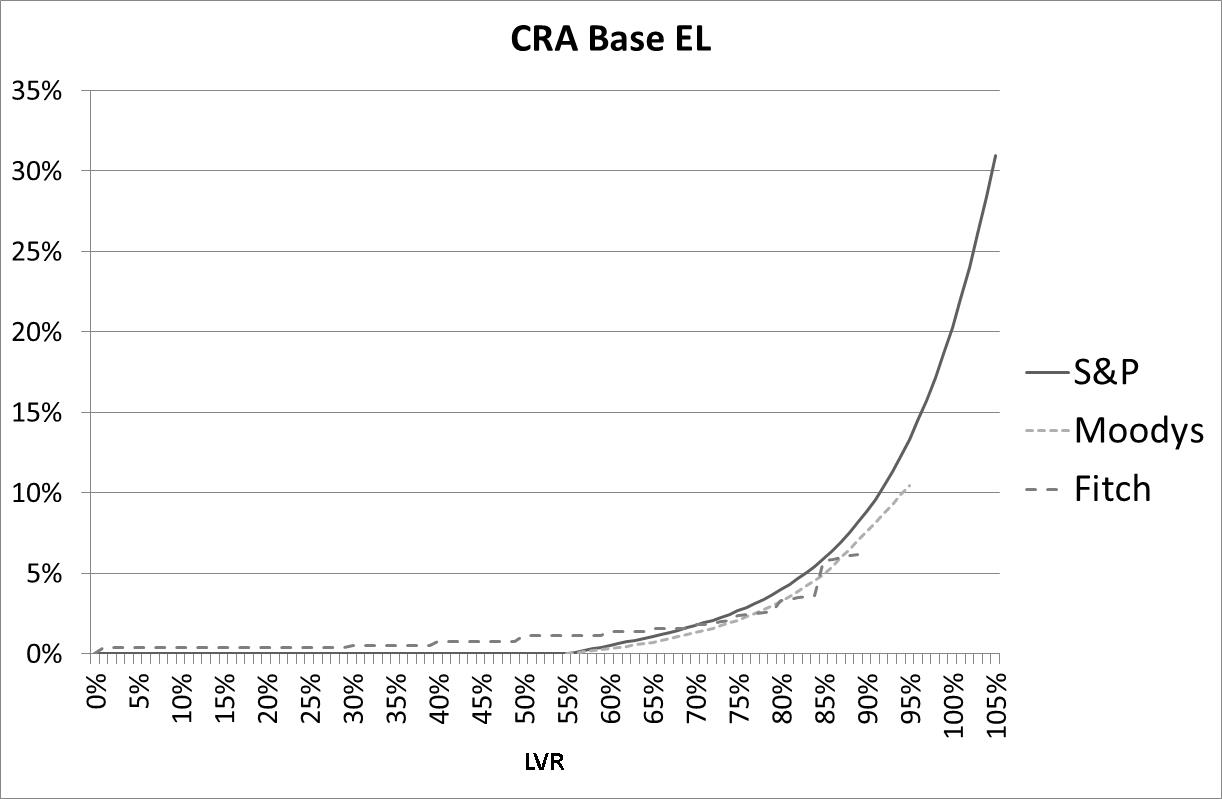}
\label{CRAEL}
\end{figure}

The CRAs refer to the product of PD and LGD variously as ‘credit enhancement’, ‘estimated loss’ and ‘estimated gross loss’.  For simplicity we refer to it as EL.  Figure \ref{CRAEL} shows EL for the CRAs.

As for PD and LGD, the different regimes appear broadly consistent with each other.
In particular, EL retains the positive curvature mentioned above for PD.

\subsection{Discussion}
As noted above, in the cases of Standard \& Poor's and Moody's, there is always an LVR below which LGD is zero.  This means that there are situations where the PD is non-zero, but LGD and EL are zero.  This clearly implies the use of an Arrears Default definition of PD, as the modelling contemplates loans which can default, but not generate any loss on Liquidation.  This is less clear in the case of Fitch (due to the non-zero minimum LGD), however the structural similarity of their model suggests a similar conclusion.

In summary, the existing approaches appear structurally inconsistent, in that they seem to adopt base PD values consistent with a Liquidation Default definition, but apply these values to generate EL in a manner consistent with an Arrears Default definition.

The authors acknowledge that this apparent structural inconsistency does not necessarily mean that the practical results produced by the various CRA criteria are inappropriate for their purposes.  In particular:
\begin{itemize}
\item The base PD, LGD and EL approaches described above are only a limited component of the overall rating process;
\item Each CRA incorporates some level of human oversight devoted to ensuring appropriate practical outcomes; and
\item The wide and long-term use of credit ratings by market participants strongly suggests that the rating agency approaches to EL are regarded by market participants as being \textit{de facto} at least reasonable.  This is further supported by the similarity of other non-public credit models.
\end{itemize}

Nevertheless we suggest that a credit risk model which treats default for Asset-Backed Loans in a structurally consistent way would be an improvement and potentially offer greater insight into credit risk.

We finally reiterate the personal experience of the authors, that other non-public approaches used by other market participants follow the same approach and caution that the foregoing section should not be interpreted as singling out the CRAs treatment as being unusual.

\section{A Consistent Mathematical Risk Framework for Asset-Backed Loans}
We now present a mathematical risk framework for Asset-backed Loans, starting from the premise that the framework should be based on the selection of a definition of default, and constructed consistently with the selected definition.  We further show that either Arrears Default or Liquidation Default can be used to create such a framework, with EL defined to be identical in each framework.

\subsection{Arrears Default Framework}
We start with the Arrears Default framework.
We first assume that PD is independent of LVR, so that for any given loan with specified characteristics other than LVR, PD is a constant, $P_a$.
For LGD, we adopt with the general approach of expressing LGD in terms of MVD.
\begin{equation}
LGD(L,M)=max\left[LDG_{min},\frac{(L-M-1)}{L}\right]
\end{equation}
As noted above, however, this definition of LGD necessarily implies a zero LGD below a certain LVR, and so necessarily implies an Arrears Default definition.  While this is not obviously problematic under an Arrears Default framework, it will prevent the development of a corresponding Liquidation Default framework.  Accordingly, we modify our expression for LGD in a way that will permit this development, while introducing a more realistic model.

The approach above treats MVD as a single-valued variable.  A lender will simply choose a value for MVD which corresponds to the level of economic stress they wish to be protected against.  Instead, we can consider that property prices in reality are volatile, and can move up or down from time to time.  On this view, MVD is not single-valued, but follows a distribution, giving the probability that the property price has migrated to any given value within the horizon of interest.

With this motivation we define MVD as not as a single parameter, but as a probability distribution where P(M) is the probability density function of the MVD.  That is, the probability of the MVD being between $M$ and $M+\delta M$ is given by:
\begin{equation}
p(M,M+\delta M)=\int^{M+\delta M}_{M}P(M)dM
\end{equation}

Specifically, $P(M)$ represents the distribution of MVD, within the default horizon of the loan, whether the loan is in arrears or default or not.

Now we can express LGD as:
\begin{equation}
LGD_a(L)=\int^{L-1}_{-1}\left(\frac{L-M-1}{L}\right)P(M)dM)
\end{equation}

Here, we are taking the single-parameter LGD, multiplying it by the probability of the corresponding MVD, and integrating from -1 (the security property loses all value) up to L-1 being the point at which the single-parameter LGD reaches zero\footnote{Here and elsewhere one could argue that the lower bound of the integral and the domain of P(M) could extend past -1, corresponding to negative effective property value, implying transaction costs greater than sales proceeds.  This issue does not have any fundamental impact on the approach.}.  It is clear that this definition is consistent with the Arrears Default definition, since it only depends on the LVR and our assumed MVD distribution.  Specifically this LGD is the expectation value of the shortfall between the current loan balance and the value of the security property, whether the loan is in arrears or default or not.

Now we can simply define EL as this LGD multiplied by $P_a$:
\begin{equation}
EL(L)=P_a\int^{L-1}_{-1}\left(\frac{L-M-1}{L}\right)P(M)dM
\end{equation}
\subsection{Liquidation Default Framework}

We can now develop a Liquidation Default framework.

To do this, we define EL as above and then ask what is the probability of a loss occurring, ie the Liquidation Default-based PD.  A loss will occur in this framework iff $L>1-M$ (ie the security property value is less than the loan balance) and there is an Arrears Default (ie the borrower is unable to meet repayment obligations).  We calculate this as follows:
\begin{equation}
P_l(L)=P_a\int^{L-1}_{-1}P(M)dM
\end{equation}

This is simply the Arrears Default PD, multiplied by the aggregate probability that $M<L-1$.

Now we can simply derive our Liquidation Default LGD as the EL divided by this PD.
\begin{equation}
LGD_l(L)=\frac{\int^{L-1}_{-1}\left(\frac{L-M-1}{L}\right)P(M)dM}{\int^{L-1}_{-1}P(M)dM}
\end{equation}

We note that where $P_l(L)$ is zero, $LGD_l$ is undefined, since its denominator is zero.  This is not simply a mathematical artefact, but represents the fact that $LGD_l$ is defined as the likely loss \textit{in the event a loss occurs}, so that its value where no loss occurs is meaningless.  We will adopt the convention of showing $LGD_l$ as zero in these cases.
\subsection{Summary of Frameworks}
In summary, we show our equation for EL, which is identical for the two frameworks, followed by PD and LGD in each framework.
\begin{equation}
EL(L)=P_a\int^{L-1}_{-1}\left(\frac{L-M-1}{L}\right)P(M)dM
\end{equation}
\begin{equation}
P_a=constant
\end{equation}
\begin{equation}
LGD_a(L)=\int^{L-1}_{-1}\left(\frac{L-M-1}{L}\right)P(M)dM)
\end{equation}
\begin{equation}
P_l(L)=P_a\int^{L-1}_{-1}P(M)dM
\end{equation}
\begin{equation}
LGD_l(L)=\frac{\int^{L-1}_{-1}\left(\frac{L-M-1}{L}\right)P(M)dM}{\int^{L-1}_{-1}P(M)dM}
\end{equation}

\section{Numerical Examples of EL derived from $P(M)$}
We present below some numerical examples of EL, PD and LGD for some simple cases of P(M).  In all examples we assume an Arrears Default probability of 7.5\%.

\subsection{Single-Value MVD}
\begin{figure}[htpb]
\caption{}
\includegraphics[width=120mm]{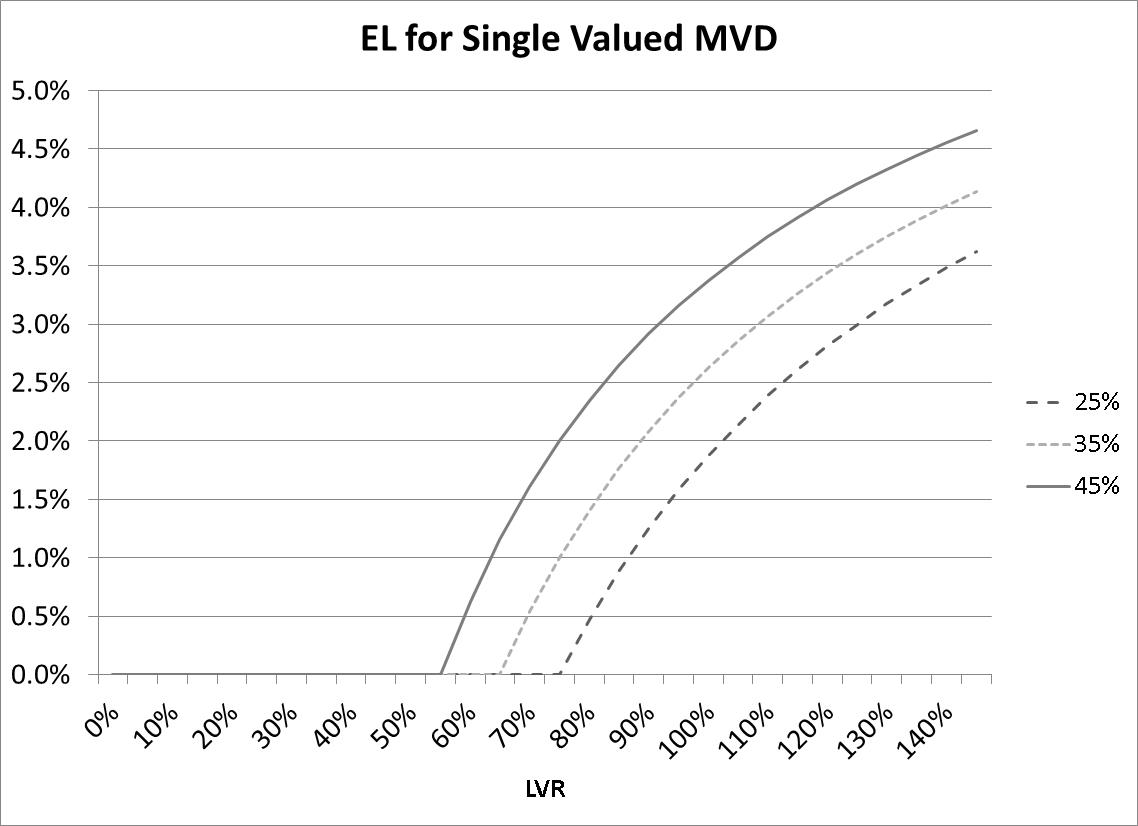}
\label{SingleMVDEL}
\end{figure}

Our first set of examples is based on the ‘trivial’ assumption of 100\% likelihood of a single-valued MVD.
\begin{equation}
P(M)=\delta (M-m)
\end{equation}
\indent where m is the single-valued MVD and $\delta$ is the Dirac delta function.
Note that this is the same assumption explicitly made under the existing CRA approach however we are applying the assumption to our new methodology.  Accordingly the results will illustrate the methodology, but should not be expected to produce useful or realistic results.
Figure \ref{SingleMVDEL} shows EL for single-valued MVD of 25\%, 35\% and 45\%.

This also gives the shape of $LGD_a$, since this is simply proportional to EL, with $P_a$ as the constant of proportionality.
It is immediately clear that these curves are not compatible with the corresponding CRA curves, since they have strongly negative curvature, rather than the positive curvature of the CRA curves.  As noted above, the de facto recognition of the reasonableness of the CRA curves strongly suggests that these single-valued MVD curves are not a realistic representation of risk.  As noted above, this is unsurprising given the assumptions used.

\begin{figure}[htpb]
\caption{}
\includegraphics[width=120mm]{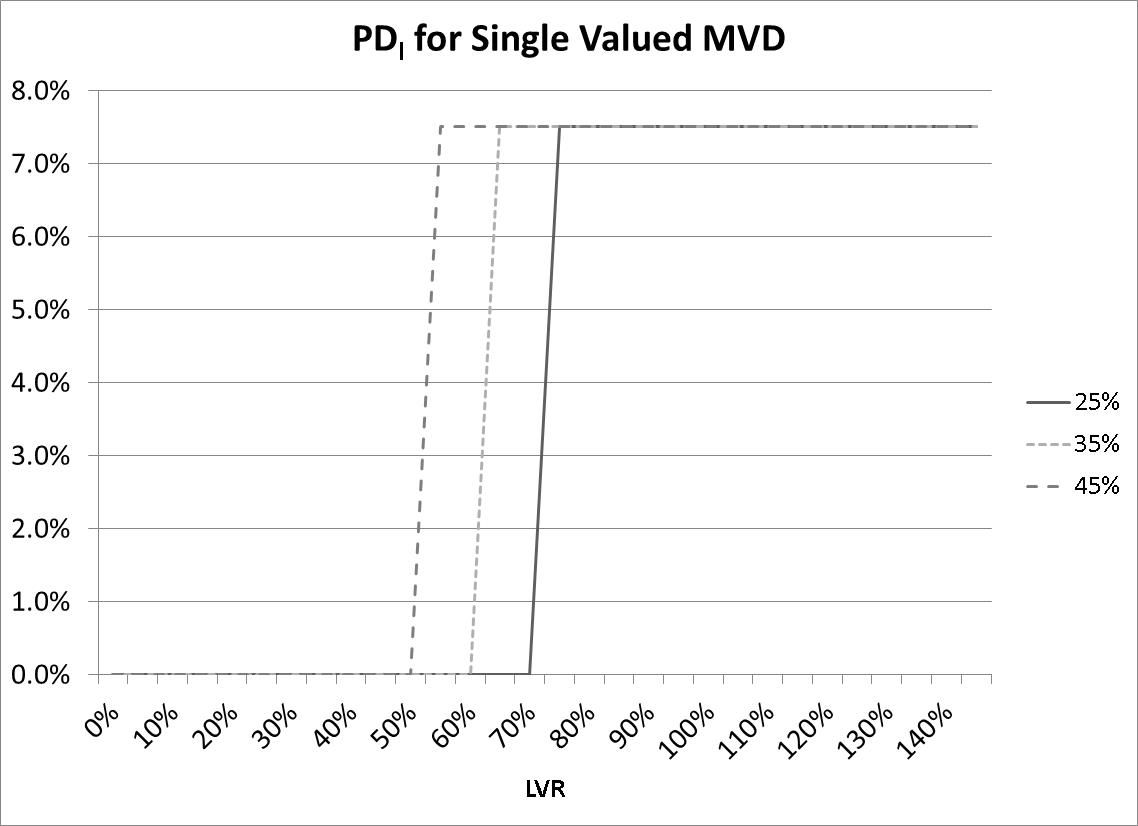}
\label{SingleMVDPD}
\end{figure}

\begin{figure}[htpb]
\caption{}
\includegraphics[width=120mm]{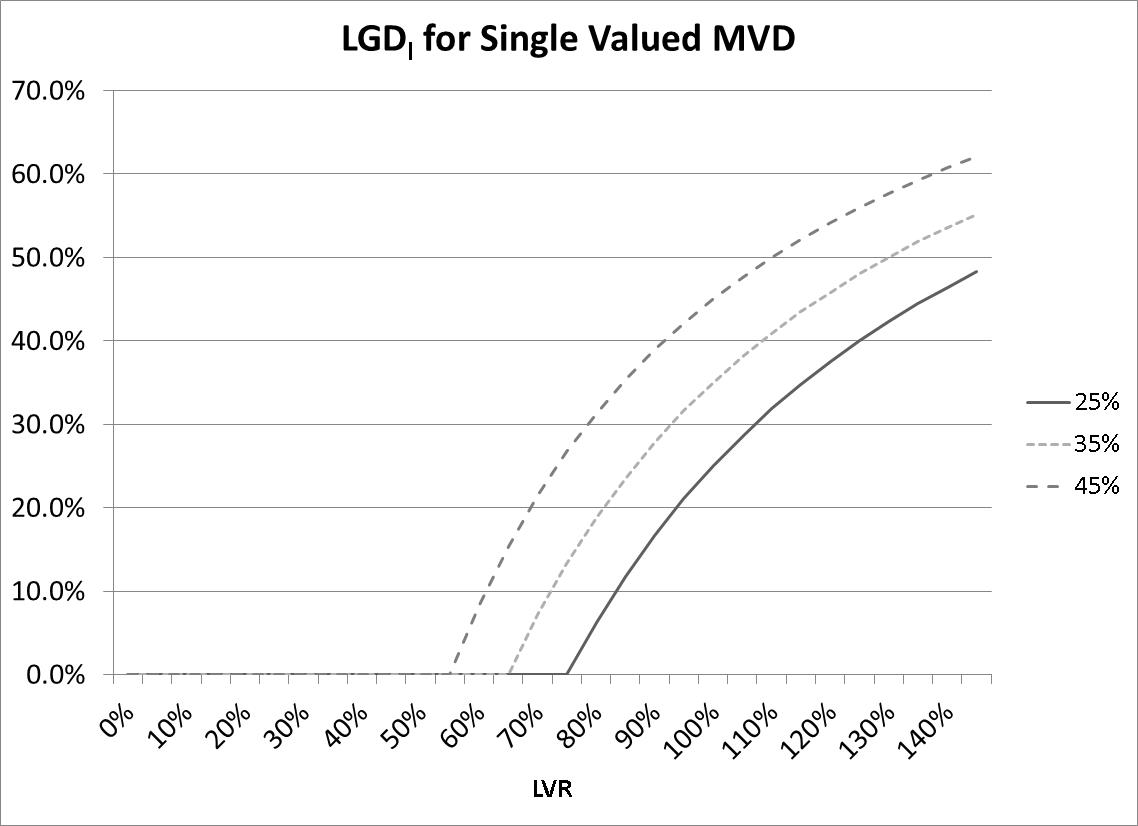}
\label{SingleMVDLGD}
\end{figure}

Figures \ref{SingleMVDPD} and \ref{SingleMVDLGD} show $PD_l$ and $LGD_l$ for the same cases.

Since $PD_l$ is just a step function, $LGD_l$ replicates the shape of the EL curve.

\subsection{Normally Distributed MVD}

Having considered single-valued MVD and found the results to be unrealistic, we now wish to consider using a more realistic distribution.
As an example, we will consider P(M) as a normal distribution.  The authors do not base this on any underlying theory or model of property value changes and acknowledge that such a model or theory may produce a very different distribution.  We further stress that our approach can be implemented with any 'reasonable' distribution.  Nevertheless, our choice of example is based on the following factors:
\begin{itemize}
\item The normal distribution is a well-understood distribution with wide application;
\item It seems intuitively reasonable to assume a central distribution; and
\item It seems intuitively reasonable to think of house prices as being affected by a large number of independent central factors, suggesting the application of the Central Limits Theorem.
\end{itemize}

Specifically, we assume that M is distributed normally around zero, with a specified standard deviation, $S_M$.

\begin{equation}
P(M)=Norm(M,M_0,S_M)\equiv \frac{1}{S_M\sqrt{2\pi}}e^{-\left(\frac{M^2}{2{S_M}^2}\right)}
\end{equation}

\begin{figure}[htpb]
\caption{}
\includegraphics[width=120mm]{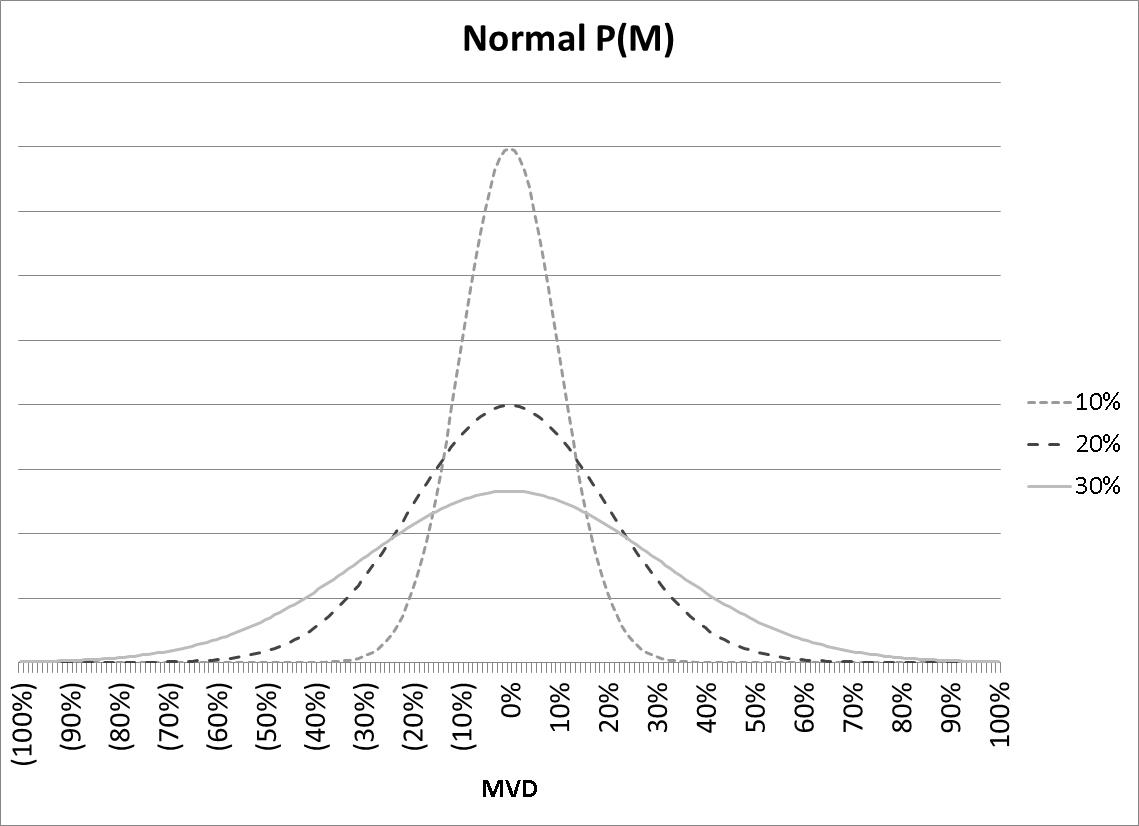}
\label{NormPM}
\end{figure}

Figure \ref{NormPM} shows P(M) for standard deviations of 10\%, 20\% and 30\%.

\begin{figure}[htpb]
\caption{}
\includegraphics[width=120mm]{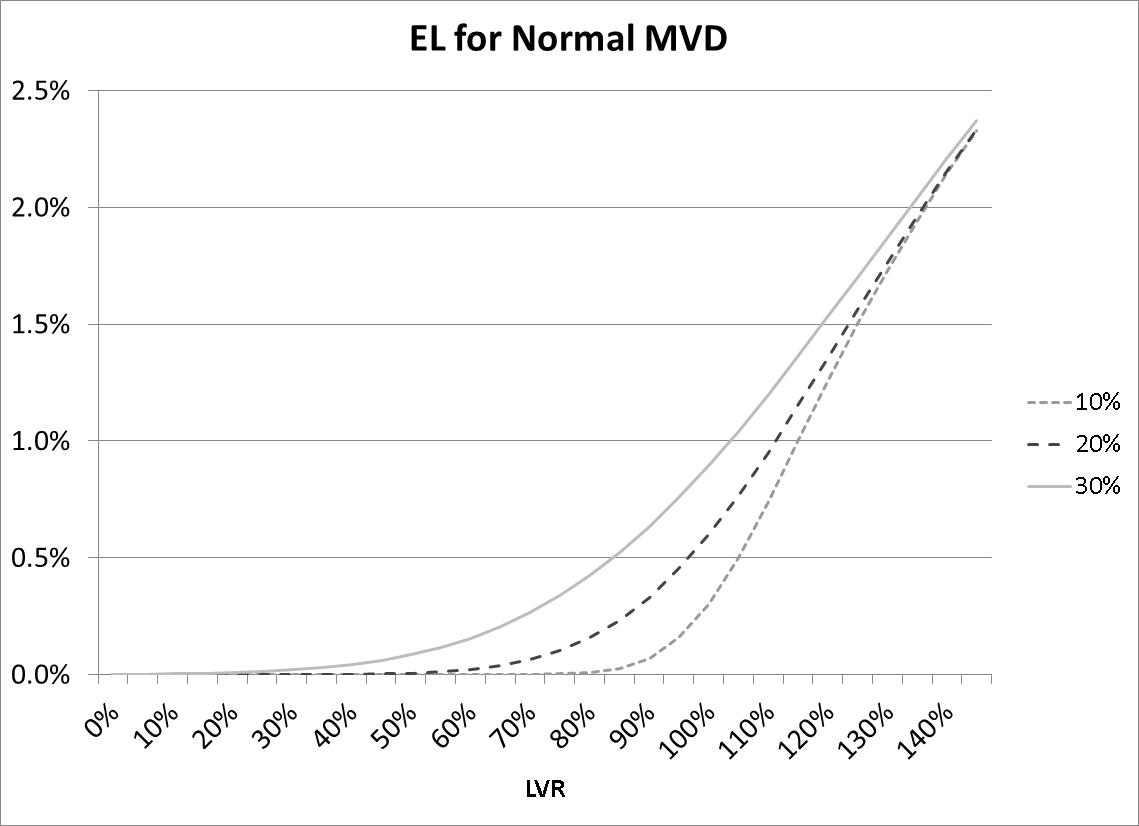}
\label{NormEL}
\end{figure}

\begin{figure}[htpb]
\caption{}
\includegraphics[width=120mm]{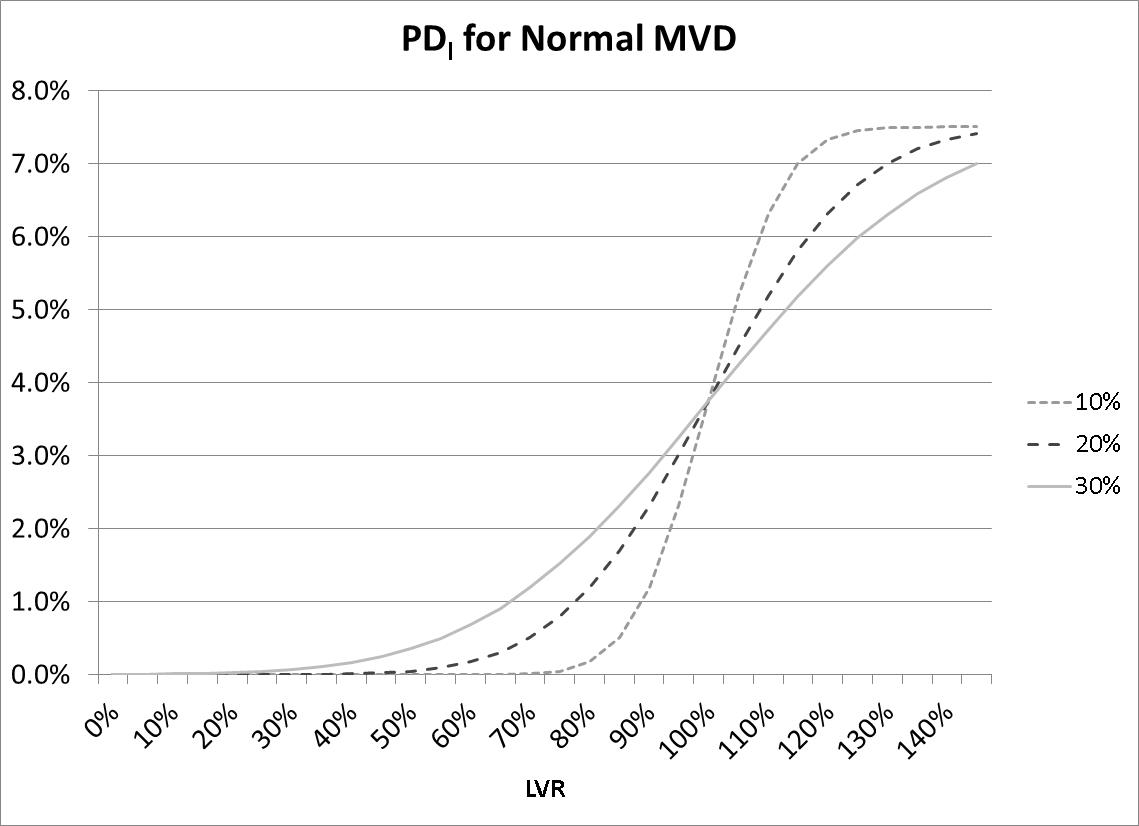}
\label{NormPD}
\end{figure}

\begin{figure}[htpb]
\caption{}
\includegraphics[width=120mm]{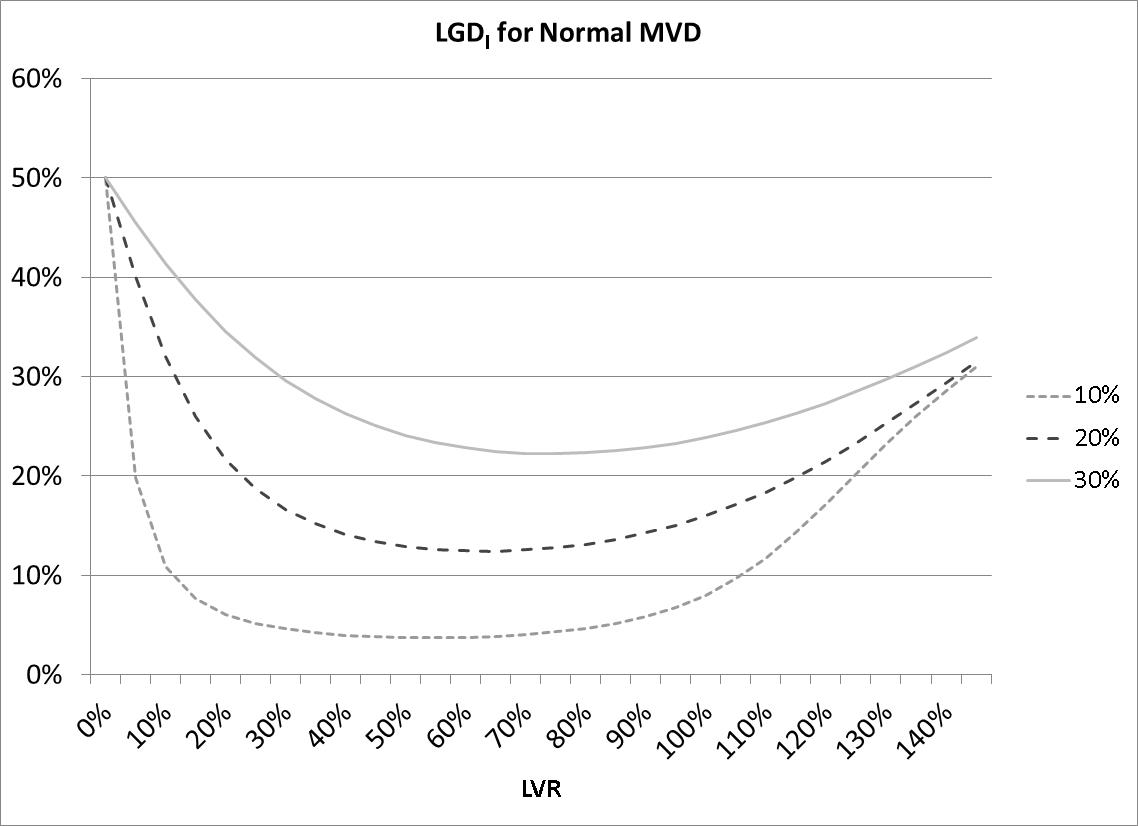}
\label{NormLGD}
\end{figure}

Figures \ref{NormEL}, \ref{NormPD} and \ref{NormLGD} show the EL,$ PD_l$ and $LGD_l$ curves for each of the above standard deviations.

The EL curves in Figure \ref{NormEL}, while not the same as the CRA curves, are more consistent.  In particularly, they show the significant positive curvature characteristic of the CRA curves in the main LVR region.  However, at very high LVR ($\sim120\%$), curvature becomes negative.  This makes intuitive sense, as in this region, we are sufficiently certain of a shortfall on Liquidation that this component of risk is no longer increasing.

Some important other features are:
\begin{itemize}
\item EL is non-zero for all positive LVR, though very small at low LVR, representing the fact that the normally distributed MVD has some small probability of even a 100\% decline;
\item The lower the MVD standard deviation, the higher the curvature of the EL curve; and
\item EL converges at very high and very low LVR.
\end{itemize}

The $PD_l$ curves in Figure \ref{NormPD} are simply cumulative normal distributions multiplied by $P_a$ (7.50\% in our example).
In line with intuition, the Liquidation Default PD depends heavily on LVR, however, contrary to the CRA approach it is capped at the constant $P_a$.  We also observe that all the curves cross at LVR of 100\%, as expected given the mean assumed MVD of 0\%.

We clarify that this should not be taken to mean that our approach has an inherent PD cap.  $P_a$ is only a constant with respect to LVR and MVD, but can vary freely based on other characteristics such as documentation type, borrower income or arrears status.  In effect, each set of characteristics affecting $P_a$ would generate a different family of the above PDl curves (whether explicitly or implicitly).

The $LDG_l$ curves in Figure \ref{NormLGD} are less intuitive, particularly as they are not monotonic with LVR.

At higher LVR, $P_l$ approaches the constant $P_a$ and LGD simply increases with LVR, approaching the shape of the EL curve.  This is intuitively sensible.

At lower LVR, EL and Pl both approach zero, however their ratio ($LGD_l$) increases. This is saying that that, at very low LVR, the likelihood of a loss becomes small very fast, but the loss incurred in the event of a loss becomes more significant.  In fact, $LGD_l$  approaches 50\% as LVR approaches zero.\footnote{Mathematically this can be seen as follows.  At very low LVR, the slope of the normal distribution approaches zero.  This means that the product of single-valued $LGD (=(L-M-1)/L)$ and P(M) approaches being proportional to the single-valued LGD.  This function is triangular from 100\% at zero LVR to 0\% at the LVR for which we are calculating LGD, so the ratio of its area to the area of (flat) P(M) approaches 50\%.}

We are not surprised at the fact that $LGD_l$ is not particularly intuitive, as it strikes us as an odd measure of risk, looking at the LGD only where a non-zero loss (however small) occurs, and ignoring the effect of cases where no loss occurs. This makes a questionable distinction between zero losses and arbitrarily small non-zero losses.

\section{Deriving P(M) from EL}
Above, we have considered the process of specifying P(M) (and $P_a$) and deriving EL and other risk measures.  It is clearly interesting to reverse this process, considering a specified EL and deriving an implied P(M) assuming a given $P_a$.

Given the form of our formula for P(M), in general there is no way of approaching this problem analytically.  Instead we must use numerical methods to derive P(M).

In realistic cases, it turns out to be a simple matter to goal-seek successive discrete values of P(M) to produce the target E(L) function.  The appendix sets out in detail the method used for the following example.

\subsection{Example of EL Derived from P(M)}
For our example, we consider a simple E(L) function designed to capture the essentials of the standard approach as set out above.  Specifically we define E(L) as follows:
\begin{equation}
EL(L)=min\left[100\%,0.015\times L^{20}\times max\left[0,\frac{L+0.4-1}{L}\right]\right]
\end{equation}
This represents:
\begin{itemize}
\item PD strongly dependent on LVR, expressed as a power function of LVR, capped at 100\%; and
\item Single-valued LGD as described above based on an assumed MVD of 40\%.
\end{itemize}

\begin{figure}[htpb]
\caption{}
\includegraphics[width=120mm]{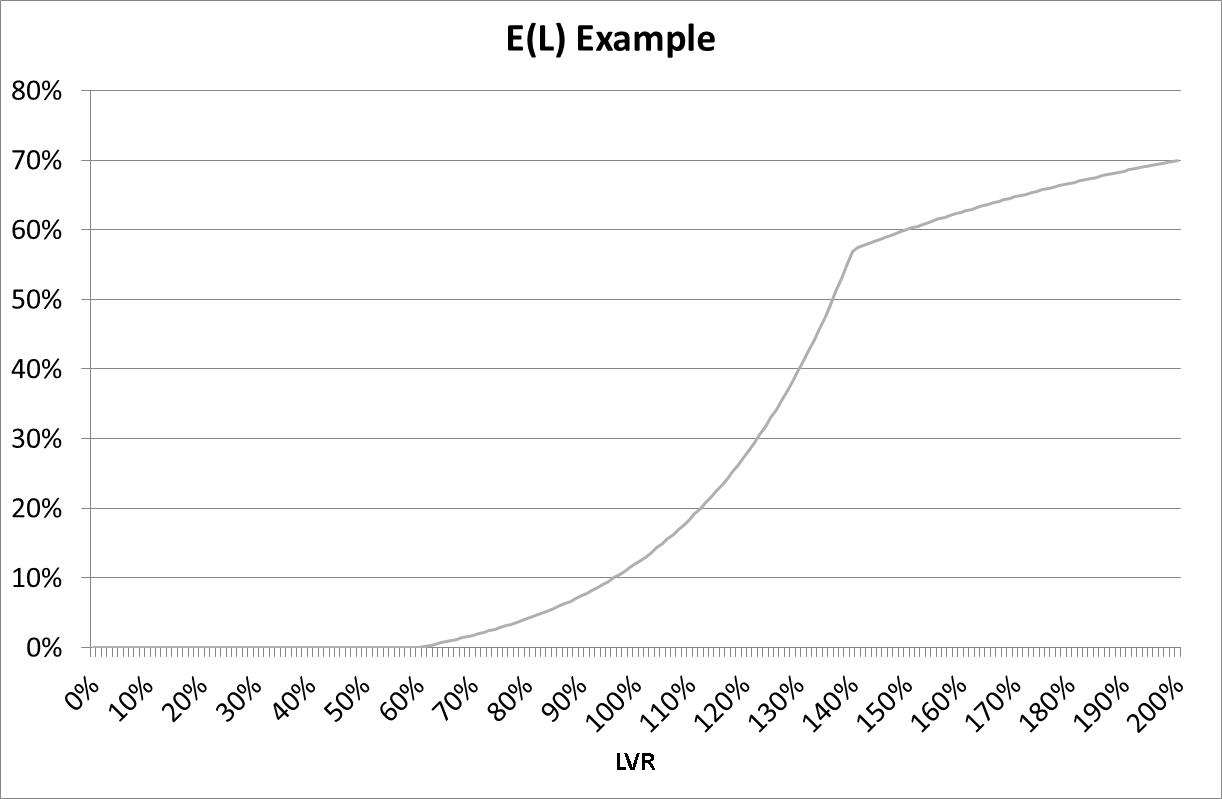}
\label{EgEL}
\end{figure}

Figure \ref{EgEL} shows the assumed EL curve.  The break at 140\% LVR is where the PD reaches 100\% and is capped.  We have assumed 1\% increments for LVR and MVD in our analysis.

\begin{figure}[htpb]
\caption{}
\includegraphics[width=120mm]{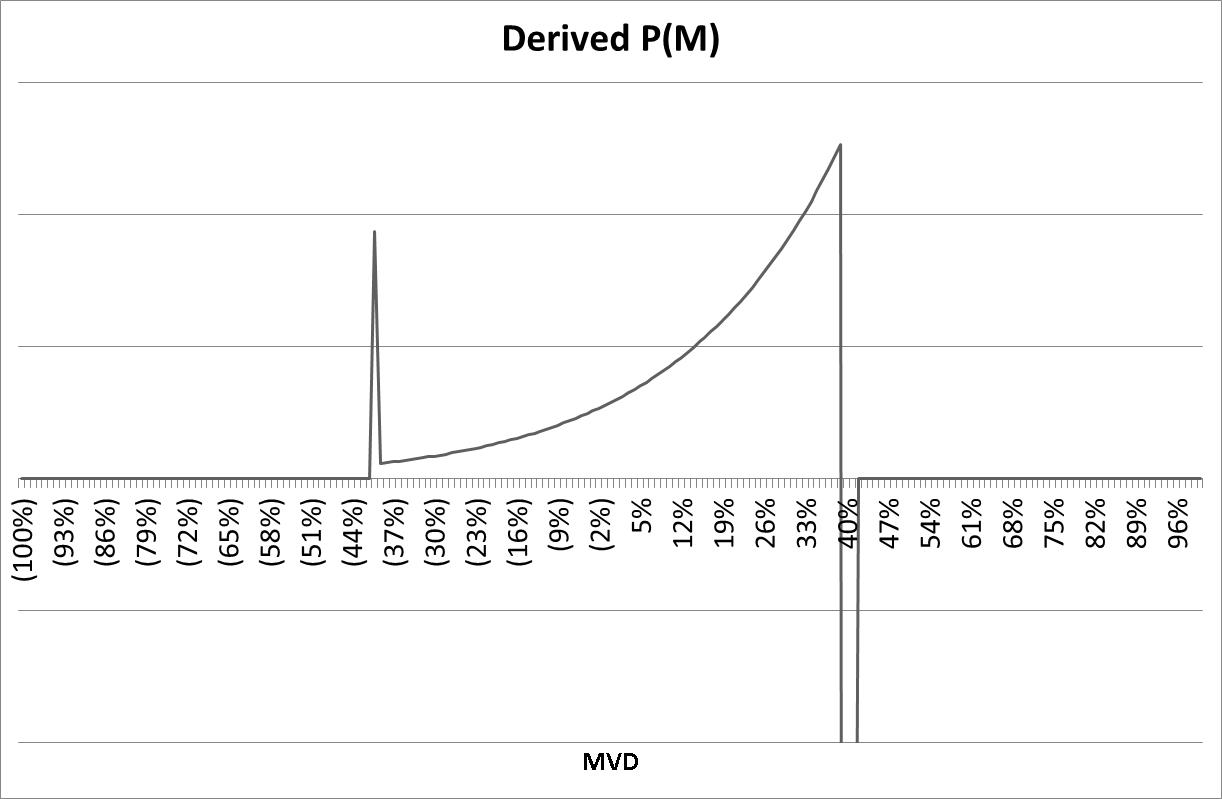}
\label{EgPM}
\end{figure}

Figure \ref{EgPM} shows the numerically derived implied MVD curve, P(M), corresponding to the EL curve above.  We have assumed $P_a$ of 10\%.

The MVD curve exhibits five distinct sections:
\begin{itemize}
\item Below a MVD of -40\%, P(M) is zero, as expected given that the assumed EL curve is based on the assumption of a -40\% MVD;
\item At -40\% MVD, there is a ‘spike’ in P(M).  This represents the fact that we have a ‘sudden’ increment in EL from zero to a non-zero value.  While for higher MVD, the increments in P(M) incorporate probability from a range of values of MVD, at this initial point, all the probability is from the single value of MVD under consideration, hence the ‘spike’;
\item From MVD -40\% to +40\%, P(M) rises smoothly, with positive curvature.  This is consistent with the underlying exponential behaviour of PD, and also seems intuitively reasonable;
\item At MVD +40\%, there is another ‘spike’, corresponding to the assumed cap in PD.  This spike, however is negative.  This corresponds to the LVR of 140\% where the assumed PD reaches its cap\footnote{Aside from the point that PD greater then 100\% is not meaningful, if we had not imposed the PD cap of 100\%, our distribution P(M) would simply continue to increase indefinitely.  Clearly this would result in a distribution that could not be normalised.};
\item Above MVD +40\%, P(M) is zero.  This corresponds to our assumed PD being fixed and capped at 100\%.
\end{itemize}

The most interesting aspect of the MVD curve is the ‘spike’ of negative probability as the assumed PD hits its cap.  Clearly negative probability is not strictly meaningful, but is a mathematical artefact produced by a discontinuity in the slope of the assumed EL curve.  Similarly, the positive spike at MVD -40\% is clearly an artefact produced by a discontinuity.  In line with comments above on existing practice, this does not necessarily mean that the assumed EL curve is problematic in practice:
\begin{itemize}
\item The spikes are limited to single points on the curve, rather than a continuous region;
\item One could argue that P(M) is not directly observable or even meaningful as a real quantity.  Rather, E(L) is the observable and meaningful curve, either through empirical observation of losses, or actual credit support requirements.  P(M) is rather the range of circumstances against which the lender wishes to be protected, or to analyse the consequences of.  On this interpretation, it is not clear that the spikes are problematic;
\item As already observed, a base EL curve is typically only a starting point for a component of a comprehensive credit analysis.  Any discussion of whether a base E(L) curve is reasonable should include a discussion of other elements of the model or process.
\end{itemize}

\begin{figure}[htpb]
\caption{}
\includegraphics[width=120mm]{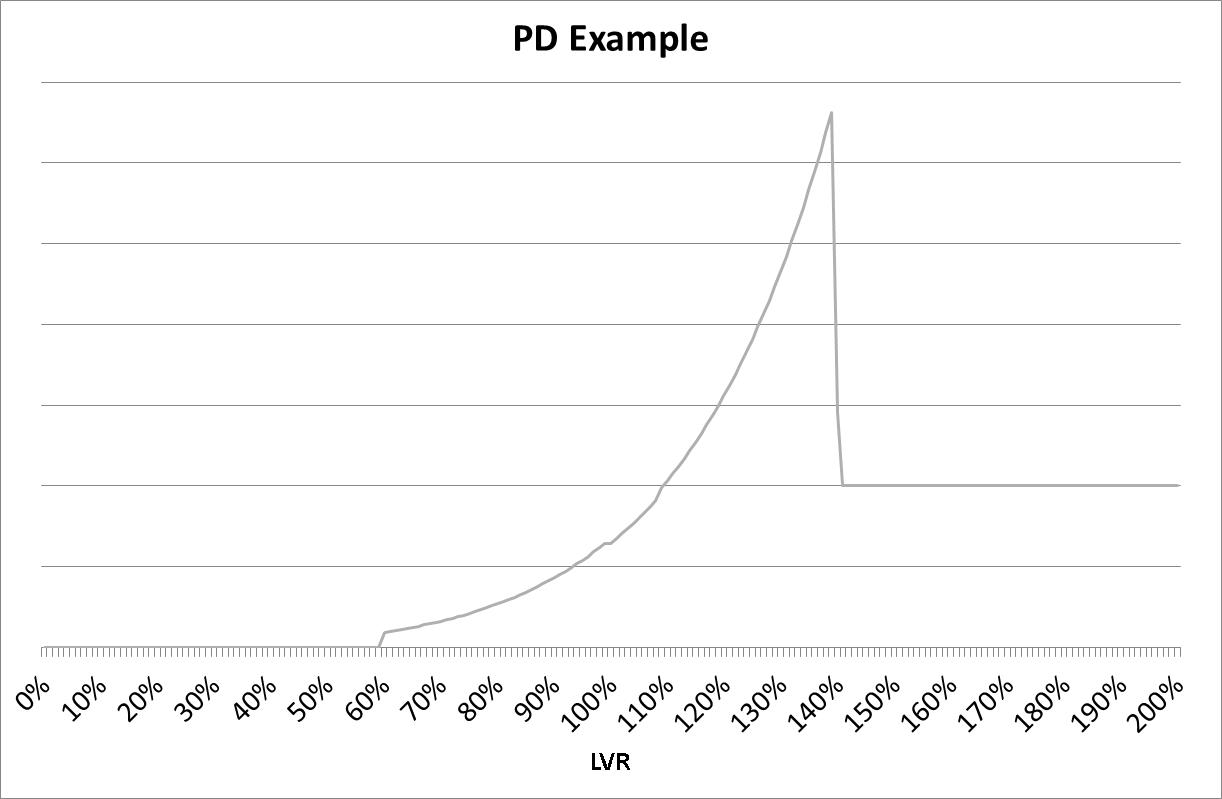}
\label{EgPD}
\end{figure}

Figure \ref{EgPD} shows $PD_l$ implied by the derived EL curve.

Clearly the section with PD $>100\%$ is unrealistic.  In line with comments above, we caution that this does not necessarily mean that the base EL curve is problematic.

Notwithstanding our comments on whether the various mathematical oddities of the implied PD and P(M) curves invalidate the given EL curve, we suggest that they highlight the internal inconsistency in current approaches to Asset-backed Loans.  A model which developed an EL curve as a representation of a consistent underlying distribution of the security asset values would provide greater insight into the drivers of risk.

\section{Extensions}
We see several obvious extensions of our methodology.
\subsection{More Refined Approaches to P(M)}
Clearly, the key driver of any E(L) curve is the choice of distribution of MVD, P(M).  While we have shown examples of normal distributions, any realistic distribution can be used.  Before any practical implementation of this approach, we suggest that a user would want to develop a customised P(M), based either on historical data, or on an economic model of property price migration.
\subsection{Varying $P_a$ with LVR}
We have assumed that $P_a$ is constant with respect to LVR.  The assumption is based on the fact that the asset value for an Asset-backed Loan has no obvious relationship to the borrowers ability to make loan repayments.

In fact, at very high LVR the borrower faces the risk that a decline in property values can result in the value of the property being significantly less than the loan balance, ‘’negative equity’.  Where a borrower is in negative equity, then there is an incentive to ‘strategically’ default on the loan.  Experian and Oliver Wyman \cite{OW} have found empirical evidence of this phenomenon.  Further, we acknowledge there may be other specific  reasons for introducing dependence on LVR. Based on this, our approach could be modified to incorporate  a specific dependence on LVR, such as increase in $P_a$ at high LVR, although we consider this dependence would bear little resemblance to the current LVR-dependent PD curves generally in use.

\subsection{Regulatory Capital}

We note that the general approach to setting regulatory capital in relation to Asset-Backed Loans under advanced methodology involves:
\begin{enumerate}
\item First determining a base or expected level of PD, assuming ‘normal’ credit conditions;
\item Then deriving a stressed PD representing a given likelihood of losses assuming log-normal distribution of losses; and
\item Finally using this stressed PD to set the required level of capital.
\end{enumerate}

Under our approach, a conceptually similar result could be obtained by simply choosing a distribution P(M) which represents an appropriate level of stress.  This could be either:
\begin{itemize}
\item A specific distribution relating to the desired level of stress; or
\item A truncation and renormalisation of an unstressed distribution, so that the probability of MVD lower than a given level was zero.  This approach would naturally lend itself to expressing the stress in centile terms, similarly to the current regulatory approach.
\end{itemize}

The advantages of this approach would seem to be:
\begin{itemize}
\item Freeing the determination of capital from the Efficient Markets Hypothesis implicit in the log-normal distribution of losses; and
\item Consolidation of the derivation of expected losses and stressed losses into the same mathematical model.
\end{itemize}

\section{Conclusion}
Our examination of the concept of PD for Asset-Backed Loans suggests that the underlying theory supporting existing approaches to credit risk is internally inconsistent.

We have proposed a consistent theoretical framework based on the concept of a probabilistic distribution of market value declines and suggest that this framework can provide greater insight into credit-risk for Asset-Backed Loans.

A clear area of further development for the framework is the practical consideration of MVD distributions.  In particular, we suggest the following would advance the framework:
\begin{itemize}
\item Analysis of historical actual data on changes in house prices; and
\item Development of a statistical economic model of how individual house prices change over time.
\end{itemize}

\appendix
\newpage
\section{Appendix – A Numerical Approach for Deriving P(M)}
We start with our expression for EL.

\begin{equation}
EL(L)=P_a\int^{L-1}_{-1}\left(\frac{L-M-1}{L}\right)P(M)dM
\end{equation}

We now convert the integral into a discrete sum.  Specifically, we divide the domain of the integral into a number of equal strips of width $\Delta M$.  This gives:

\begin{equation}
EL(L)=P_a\sum^{i_{max}}_{i=1}\left(\frac{L-M_i-1}{L}\right)P(M_i)\Delta M
\end{equation}

where $i_{max}$ is the number of strips and $M_i$ runs from -1 to L-1.  The lowest possible value of L is zero, where our formula is undefined but we define both LGD and EL to be zero.

We now use this relation to take a given EL(L) and derive the corresponding implied P(M)

First consider a single strip, moving from $EL(0)$ to $EL(\Delta L)$, where $\Delta L=\Delta M$ and $i_{max}=1$.

\begin{equation}
EL(\Delta L)=P_a\sum^{1}_{i=1}\left(\frac{\Delta L-M_i-1}{\Delta L}\right)P(M_i)\Delta M=P_a\left(\frac{\Delta L-M_i-1}{\Delta L}\right)P(M_i)\Delta M
\end{equation}

Clearly it is a trivial matter to find a satisfactory value of P(M) for this single strip, producing the desired value of $EL(\Delta L)$.

Next, we iterate this process, increasing the number of strips by 1 each time.  The k’th step is given by:

\begin{equation}
EL(L_{k-1}+\Delta L)=P_a\sum^k_{i=1}\left(\frac{L_{k-1}+\Delta L-M_k-1}{L}\right)P(M_k)\Delta M
\end{equation}

Clearly E(L) only depends on P(M) for the range M<L-1.  Equivalently, at each value of k, E(L) only depends on $P(M_i)$ for values up to k.  This means that if we have found (by whatever numerical method) appropriate values of P(M) for a particular LVR and imax, then when we progress to the next higher value, k, and try to derive P(M) for $LVR=L+\Delta L$,  we are guaranteed that only the value of P(M) for k is relevant.

The practical consequence is that we may adopt the following process:
\begin{enumerate}
\item First goal-seek a value of $P(-1+\Delta M)$ producing $EL(\Delta L)$;
\item Next goal-seek a value of $P(-1+2\Delta M)$ producing $EL(2\Delta L)$, given the already derived value of $P(-1+\Delta M)$;
\item Continue iterating – at each step, goal-seek a value of $P(-1+k\Delta M)$ producing $EL(k\Delta L)$, given the already derived values of $P(M)$ for lower k.
\end{enumerate}

\end{document}